\documentstyle[12pt]{article}

\def\be{\begin{equation}}
\def\ee{\end{equation}}
\def\bea{\begin{eqnarray}}
\def\eea{\end{eqnarray}}

\def\det{{\rm det}}
\def\haf{\frac{1}{2}}
\def\nn{\nonumber\\}
\def\tr{{\rm Tr}}
\def\e{{\rm e}}

\begin{document}

\begin{flushright}
hep-th/9808046\\
IPM-98 
\end{flushright}

\pagestyle{plain} 

\begin{center}
{\Large {\bf Interaction of Branes at Angles in M(atrix) Model}\\ }
\vspace{.5cm}
 
Kamran Kaviani$_{a,b}$, Shahrokh Parvizi$_{a}$, Amir H. Fatollahi$_{a,c}$

\vspace{.2 cm} 
{\it a)Institute for Studies in Theoretical Physics and Mathematics 
(IPM),}\\ {\it P.O.Box 19395-5531, Tehran, Iran}\\ 

\vspace{.2 cm} 
{\it b)Department of Physics, Az-zahra University,\\
P.O.Box 19834, Tehran, Iran}\\

\vspace{.2 cm} 
{\it c)Department of Physics, Sharif University of Technology,\\ 
P.O.Box 11365-9161, Tehran, Iran}\\ 
\vspace{.2 cm} 

{\sl E-mails: kaviani, parvizi, fath@theory.ipm.ac.ir}

\vspace{.2 cm} 
\begin{abstract} 
Interactions of relatively rotated Dp-branes in 1, 2, 3 and 4 angles in
M(atrix) model are calculated and it is found to be in agreement with
string theory calculations. In 4 angles case the agreement is achieved
after subtracting the contribution of the single chiral fermionic zero
mode. 
\end{abstract} 
\end{center}

M(atrix) model as a non-perturbative formulation of M-theory theory 
has activated a great effort in studies for finding the unified string
theory. The model is the dimensional reduction of the 9+1 
dimensional ${\cal N}=1$, $U(N)$ SYM to 0+1 dimension. 
This theory has passed several checks \cite{ToBa}.

The aim of this letter is the calculation of interactions of rotated 
branes by the model and comparing them with the known results to 
provide a further test for this model. In summary we have found that
the result of M(atrix) model calculation is in agreement with similar
ones in string theory. In the more novel case of rotated 4-branes 
with four angles it is found that by removing the chiral fermionic 
zero mode in defining the determinant one can recover the contribution
 $R(-1)^F$ of string theory to get the correct result.

Rotated D-branes are important due to their role in constructing 
configurations with different fraction of preserved supersymmetry
\cite{SJ,TI}. Also in the 4 angles case they have been considered in the 
context of anomalous creation of fundamental strings 
\cite{HoWu, Ohta-shi-zhou, kitao-ohta-zhou}.

The BFSS Lagrangian in units $2\pi\alpha'=1$ is \cite{BFSS}
\bea\label{2.1}
L=\frac{1}{2g_s} \tr \left(D_t X_iD_t X_i+i \theta^T D_t \theta +\frac{1}{2}
[X_i,X_j]^2+\theta^T\gamma_i[\theta,X_i]\right),\nonumber\\
D_t *=\partial_t *-i[A_0,*],\;\;\;i,j=1,...,9,
\eea
where $X_i$ and $\theta$ are bosonic and fermionic $SO(9)$ 
hermitian $n\times n$ matrices.
The static classical equations of motion with $A_0=\theta=0$ are 
\bea\label{2.5}
\sum_i [X^i,[X^i,X^j]]=0. 
\eea
Every configuration with $[X^i,X^j]\sim 1$ and 
with the other $X$'s vanishing are solutions of (\ref{2.5}).

Solutions which can be interpreted as D$p$-brane have the form 
\be 
X_i^{cl} = \left(B_1,B_2,\ldots,B_p,0,\ldots,0 \right), 
~~~~~A^{cl}_0=\theta=0\,, 
\ee
where $B_1,\ldots,B_p$ are $n\times n$ matrices with $n$ large, with
the commutation relations  
\be 
[B_a,B_b]=ic_{ab}{\bf 1},\;\;\;a,b=1,2,\dots,p.
\ee
            
By a proper rotation the anti--symmetric matrix $c_{ab}$ can be
brought to the Jordan form 
\bea 
c_{ab}=\left( 
\begin{array}{ccccc}
0 & \omega_1 &  &  &  \\ 
-\omega_1 & 0 &  &  &  \\  
&  & \ddots &  &  \\  
&  &  & 0 & \omega_l \\  
&  &  & -\omega_l & 0 \\  
\end{array}
\right), 
\eea
where $2l\equiv p=2,4,6,8$.

These solutions can be represented by 
\bea
\left\{ 
\begin{array}{lll}
X_{2i-1}&=&1_{n_1}\otimes 1_{n_2}\otimes ...\otimes 
\frac{L_{2i-1}}{\sqrt{2\pi n_i}}q_i
\otimes 1_{n_{i+1}}\otimes...\otimes 1_{n_l},\\ 
X_{2i}&=&1_{n_1}\otimes 1_{n_2}\otimes ...\otimes 
\frac{L_{2i}}{\sqrt{2\pi n_i}}p_i\otimes 1_{n_{i+1}}\otimes...
\otimes 1_{n_l}, \;\;\;\;\;\;l\geq i \\  
X^i &=&0,\;\;\;\;\;\;\;\;\;\; i > 2l=p,   
\end{array}
\right. 
\eea
\noindent where $n_1 n_2...n_l=n$ and $L_a$'s are compactification radii,
with commutation relations 
$$
[ q_{i}, p_{j} ]= i\delta_{ij}\; 1_{n_i}.
$$
The eigenvalues of $q,\;p$ are uniformly distributed as 
$$
-\sqrt{\frac{\pi n_i}{2}}\leq q_i, p_i \leq \sqrt{\frac{\pi n_i}{2}}. 
$$
So the extension of solutions along $X_i$ axis is 
$L_i\rightarrow \infty$. 
Thus one can obtain 
\bea\label{2.10}
[X_{2i-1}, X_{2i}]= {\frac{i}{2\pi n_i}} L_{2i-1} L_{2i} 1_n,
\;\;\;\;0\leq i\leq l,
\eea
and correspondingly 
$\frac{n^\frac{1}{l}}{L_{2i-1}L_{2i}} = \frac{1}{2\pi\omega_i}$
by $n_i\sim n^\frac{1}{l}$ \cite{BSS}.

The configurations with four rotation angles can be obtained from the
block-diagonal matrices with two identical blocks describing a pair of
D$p$-branes. Translating along the $(p+5)$-th axis by the distance $
r $ from each other and rotating in opposite directions through the 
angles $ \psi_4 /2,\; \psi_3 /2,\;\psi_2 /2$ and $ \psi_1 /2$, we obtain the
configuration of two rotated D$p$-branes 
\bea\label{2.35}
 X_{a}^{cl}&=&\left(
 \begin{array}{cc}
 B_a & 0 \\
 0 & B_a \\
 \end{array}
 \right),~~~~a=1,\ldots,p-4,\nonumber \\ 
 X_{p-3}^{cl}&=&\left( 
 \begin{array}{cc}
 B_{p-3}\cos\frac{\psi_4}{2} & 0 \\
 0 & B_{p-3}\cos\frac{\psi_4}{2} \\
 \end{array}
  \right), \;\;\;\;     
 X_{p-2}^{cl}=\left( 
 \begin{array}{cc}
 B_{p-2}\cos\frac{\psi_3 }{2} & 0 \\
 0 & B_{p-2}\cos\frac{\psi_3 }{2} \\
 \end{array}
  \right) , \nonumber \\  
 X_{p-1}^{cl}&=&\left( 
 \begin{array}{cc}
 B_{p-1}\cos\frac{\psi_2 }{2} & 0 \\
 0 & B_{p-1}\cos\frac{\psi_2 }{2} \\
 \end{array}
 \right),  \;      \;\;\;
 X_{p}^{cl}=\left(
 \begin{array}{cc}
 B_{p}\cos\frac{\psi_1 }{2} & 0 \\
 0 & B_{p}\cos\frac{\psi_1 }{2} \\
 \end{array}
 \right),  \nonumber \\ 
 X_{p+1}^{cl}&=&\left(
 \begin{array}{cc}
 B_{p}\sin\frac{\psi_1}{2} & 0 \\
 0 & -B_{p}\sin\frac{\psi_1}{2}\\
 \end{array}
 \right),  \;      \;\;\;  
 X_{p+2}^{cl}=\left(
 \begin{array}{cc}
 B_{p-1}\sin\frac{\psi_2 }{2} & 0 \\
 0 & -B_{p-1}\sin\frac{\psi_2 }{2} \\
 \end{array}
 \right),  \nonumber \\
 X_{p+3}^{cl}&=&\left( 
 \begin{array}{cc}
 B_{p-2}\sin\frac{\psi_3 }{2} & 0 \\
 0 & -B_{p-2}\sin\frac{\psi_3 }{2} \\
 \end{array}
  \right)  ,  \;   \;\;\; 
 X_{p+4}^{cl}=\left( 
 \begin{array}{cc}
 B_{p-3}\sin\frac{\psi_4 }{2} & 0 \\
 0 & -B_{p-3}\sin\frac{\psi_4 }{2} \\
 \end{array}
  \right)   ,   \nonumber \\  
 X_{p+5}^{cl}&=&\left(
 \begin{array}{cc}
 \frac{r}{2} & 0 \\
 0 & -\frac{r}{2} \\
 \end{array}
 \right),     \;\;\;\;
 A_{0}^{cl}=X_{i }^{cl}=0,~~~~i =p+6,\ldots,9 \,.
 \end{eqnarray}

To calculate the one-loop effective action it is convenient 
to work with a compact form of (1) after the Wick rotation 
$t\rightarrow it$ and $A_0 \rightarrow -iA_0$  
\bea\label{2.20}
L=\frac{1}{g_s}\tr\left(\frac{1}{4}[X_\mu,X_\nu]^2
+\frac{1}{2}\theta^T\gamma^\mu[X_\mu,\theta]\right),\nonumber\\
\mu ,\nu =0,1,...,9,\;\;\;X_0=i \partial _t +A_0,
\eea
where sums are with Euclidean metric and $\gamma_0=-i$.

The one-loop effective action $W$ with the backgrounds $\theta=A_{0}^{cl}=0$
was calculated in \cite{FT,ChT}. 
In our notation the bosonic, fermionic and ghost contribution
can be written as, 
\bea\label{2.30}  
W=-\ln\bigg(
\det^{-\haf}(P_\lambda^2\delta_{\mu\nu}-2iF_{\mu\nu})\cdot 
\det^{\haf}(\partial_t+\sum_{i=1}^{9}P_i\gamma_i)\cdot
\det(P_\lambda^2)\bigg),
\eea
with $P_\nu*=[X_\mu^{cl},*]$, $F_{\mu\nu}\;*=[f_{\mu\nu},*]$, 
$f_{\mu\nu}=i[X_\mu^{cl},X_\nu^{cl}]$,
\bea
P_\lambda^2=-\partial_t^2+\sum_{i=1}^9 P_i^2
\;\;{\rm and}\;\;\;\;F_{0i}=0,
\eea
which the last equality is true for static configurations with 
$A^{cl}_0=0$.

It is worth mentioning that the configurations with $F_{\mu\nu}\equiv 0$
for all $\mu,\;\nu$ have vanishing quantum corrections due to the algebra
\bea\label{2.32}
W\sim Trlog(P_\lambda^2)\;(\frac{10}{2}- \frac{16}{4}- 1)=0.
\eea

By setting $\omega_1=\omega_2=...=\omega$ one finds 
\bea\label{2.40}
f_{2a-1,2a}&=&-\omega\otimes 1,\;\;a=1,...,l-2,\nonumber\\
f_{p-3,p-2}&=&-\omega\cos\frac{\psi_3}{2}\cos\frac{\psi_4}{2}\otimes 1,
\;\;\;\;
f_{p-3,p+3}=-\omega\cos\frac{\psi_4}{2}\sin\frac{\psi_3}{2}\otimes
\sigma_3,\nonumber\\
f_{p-2,p+4}&=&\omega\cos\frac{\psi_3}{2}\sin\frac{\psi_4}{2}\otimes
\sigma_3,\;\;\;\;
f_{p+3,p+4}=\omega\sin\frac{\psi_3}{2}\sin\frac{\psi_4}{2} 
\otimes 1,\nonumber\\
f_{p-1,p}&=&-\omega\cos\frac{\psi_2}{2}\;\cos\frac{\psi_1}{2}\otimes 1,
\;\;\;\;\;
f_{p-1,p+1}=-\omega\cos\frac{\psi_2}{2}\;\sin\frac{\psi_1}{2}\otimes\sigma_3,
\nonumber\\ 
f_{p,p+2}&=&\omega\cos\frac{\psi_1}{2}\;\sin\frac{\psi_2}{2}\otimes\sigma_3,
\;\;\;\;\;
f_{p+1,p+2}=\omega\sin\frac{\psi_1}{2}\;\sin\frac{\psi_2}{2}\otimes 1,
\nonumber\\
&~&{\rm otherwise}\;\;\;f_{ab}=0,
\eea
where $\sigma_3$ is the Pauli matrix. Then
\bea\label{2.45}
[P_{p-3},P_{p+3}]&=&i\omega\cos\frac{\psi_4}{2}\sin\frac{\psi_3}{2}
\otimes\Sigma_3,    \;\;\;\;
[P_{p-2},P_{p+4}]=-i\omega\cos\frac{\psi_3}{2}\sin\frac{\psi_4}{2}
\otimes\Sigma_3,\nonumber\\
{}[P_{p-1},P_{p+1}]&=&i\omega\cos\frac{\psi_2}{2}\sin\frac{\psi_1}{2}
\otimes\Sigma_3,  \;\;\;\;\;\;
[P_{p},P_{p+2}]=-i\omega\cos\frac{\psi_1}{2}\sin\frac{\psi_2}{2}
\otimes\Sigma_3,\nonumber\\
&~&{\rm otherwise}\;\;\;F_{ab}=0, \;\;\; P_{p+5}=\frac{r}{2}\otimes\Sigma_3,
\eea
where $\Sigma_3 *=[1\otimes\sigma_3,*]$. $\Sigma_3$ has 2, -2, 0 and 0 as 
eigenvalues. 

We define the following parameters for later conventions:
\bea
U_{1}&=&2\omega\cos{\frac{\psi_4}{2}}\sin{\frac{\psi_3}{2}},
\;\;\;\;\;
U_{2}=2\omega\cos{\frac{\psi_3}{2}}\sin{\frac{\psi_4}{2}},
\nonumber\\
U_{3}&=&2\omega\cos{\frac{\psi_2}{2}}\sin{\frac{\psi_1}{2}},
\;\;\;\;\;
U_{4}=2\omega\cos{\frac{\psi_1}{2}}\sin{\frac{\psi_2}{2}}.
\eea
So one finds for (\ref{2.30})
\bea
W=-\ln&\bigg(&
\det^{-1}(P_\lambda^2)\nn
&~&\times\det^{-\haf}(P_\lambda^2-2U_1)
  \det^{-\haf}(P_\lambda^2+2U_1)\nn
&~&\times\det^{-\haf}(P_\lambda^2-2U_2)
  \det^{-\haf}(P_\lambda^2+2U_2)\nn
&~&\times\det^{-\haf}(P_\lambda^2-2U_3)
  \det^{-\haf}(P_\lambda^2+2U_3)\nn
&~&\times\det^{-\haf}(P_\lambda^2-2U_4)
  \det^{-\haf}(P_\lambda^2+2U_4)\nn
&~&\times\det(P_\lambda^2)\;
\det^{\haf}(\partial_t+\sum_{i=1}^{9}P_i\gamma_i) \bigg).
\eea

In the cases of our interest we have $[X_\mu^{cl}, f_{\mu\nu}]=c-number$, and 
so $P_\lambda^2$ and $F_{\mu\nu}$ are simultaneously diagonalisable.

{\bf 1) Four angles} 

The four angles case is allowed only for 4-branes, $p=4=2l$.
Also in the four angles case the second term in 
(\ref{2.30}) can not bring to the Klein-Gordon form 
because of the absence tenth $16\times 16$ matrix
which anti-commutes with all the $\gamma_i$'s. But an expansion for small
$r$ is possible and one finds
\bea\label{2.55}
&~&-\ln\det^\haf(\partial_t+\sum_{i=1}^{9}P_i\gamma_i)=
-\haf \tr \ln (\partial_t+\sum_{i=1}^{8}P_i\gamma_i+P_9\gamma_9)=
\nonumber\\
&~&-\frac{1}{4}\tr\ln(P_\lambda^2+\frac{i}{2}F_{ij}\gamma_{ij})-\frac{r}{2}
\tr(\frac{1}{\partial_t\gamma_9+\sum_{i=1}^{8}P_i\gamma_9\gamma_i})+ O(r^2),
\eea
which the last trace is shown to vanish in the Appendix.

The eigenvalues of $P^2_\lambda$ can be read from (\ref{2.45}).
Zero eigenvalues of $\Sigma_3$ will not have any contribution 
to the effective action because of (\ref{2.32}). The other two 
eigenvalues force $(P_{p-3},P_{p+3})$, $(P_{p-2},P_{p+4})$, $(P_{p-1}
,P_{p+1})$ and $(P_{p},P_{p+2})$ to behave as harmonic oscillators 
with their related frequencies to be read from (\ref{2.45}).
The eigenvalues of $P_\lambda^2$ are 
\bea\label{2.60} 
E=r^2+q_0^2 +2\sum_{a=1}^4 U_a(k_a+\frac{1}{2}),
\eea
with $q_0$ as eigenvalues of $i\partial_t$, and $k_a$ as
harmonic oscillator numbers.

Using Schwinger's time representation, one finds the 
one-loop effective action $W$ as,
\bea
W=\int dt\int_{0}^{\infty}\frac{ds}{s} \e^{-s(q_0^2+r^2)}
 \sum_{k_1,\cdots,k_4=0}^{\infty} \e^{-s\sum_{a=1}^{4} U_a(2k_a+1)} 
\nn
 \times
\left[
2\sum_{a=1}^4  (\e^{-2sU_a}+\e^{2sU_a})
-\sum_{{\bf s}_1,\ldots,{\bf s}_4=\pm 1}
 \e^{-s\sum_{a=1}^4 U_a{\bf s}_a}
\right],
\eea
where ${\bf s}_{1,2,3,4}$ are the eigenvalues of the commutative matrices
$i\gamma_{17},\;i\gamma_{82},\;i\gamma_{35},\;i\gamma_{64}$ respectively
and the terms in the last bracket come from bosons and fermions 
respectively.
In the above expression  for $W$ one can specify the contribution of
a chiral fermionic zero mode defined by 
$$
\left( \sum_{i=1}^8 P_i\gamma_i \right)\theta_0 =0,
$$ 
for $k_1=k_2=k_3=k_4=0$ and 
${\bf s}_1={\bf s}_2={\bf s}_3={\bf s}_4=-1$ in (19)
\footnote{By chirality we mean in the 8 dimensional subspace defined 
by the two 4-branes as
$$
(i\gamma_{17})(i\gamma_{82})(i\gamma_{35})(i\gamma_{64})\theta_0
=(-1)^4\theta_0
=\gamma_1\gamma_2\gamma_3\gamma_4\gamma_5\gamma_6\gamma_7\gamma_8\theta_0
=\gamma_9\theta_0.
$$}
\cite{HoWu,Li}.

To calculate the one-loop effective action one should project out the 
contribution of the chiral fermionic zero mode \cite{Vech, Negele}
to be done here as
\footnote{The same is true for D0-D8-brane system studied in 
\cite{Pie}.}
\bea
W'=W-\int dt\int_{0}^{\infty}\frac{ds}{s} \e^{-s(q_0^2+r^2)} (-1).
\eea
By putting $\omega<<1$ one finds,
\bea
W'&=& \int dt \int_{0}^\infty\frac{ds}{s}
\left( \frac{\pi}{s} \right)^\frac{1}{2} \e^{-sr^2}
\nn
&~&\times\frac{
\sum_{a=1}^4\cos(2\psi_a)-4\prod_{a=1}^4\cos\psi_a
+4\prod_{a=1}^4\sin\psi_a+{\cal{O}}(s^2)}
{4\prod_{a=1}^4\sin\psi_a},   
\eea
and finally
\bea
W'= -2\sqrt{\pi}\int dt 
\frac{
\sum_{a=1}^4\cos(2\psi_a)-4\prod_{a=1}^4\cos\psi_a
+4\prod_{a=1}^4\sin\psi_a}
{4\prod_{a=1}^4\sin\psi_a}   |r| ,
\eea
in agreement with string theory calculations \cite{SJ}.

The above interaction vanishes in 
$\psi_1+\psi_2+\psi_3+\psi_4=0\;(mod\;2\pi)$ 
cases, signalling enhancement of SUSY. An equivalent result is obtained in
\cite{SJ} by considering interactions of rotated D$p$-branes, 
and in \cite{TI} by studying the SUSY algebra for rotated M-objects.


{\bf 2) Three angles}

By putting $\psi_4=0$ in (8) one finds the relevant 
configuration and commutation relations for the 3 angles case.
Again the 3 angles are accessible only for $p=4$.
                                  
The eigenvalues of $P^2_\lambda$ can be read from (\ref{2.45}).
Zero eigenvalues of $\Sigma_3$ will not have any contribution 
to the effective action because of (\ref{2.32}). 
As the four angles case one has harmonic oscillators 
with their related frequencies to be read from (\ref{2.45}).
The eigenvalues of $P_\lambda^2$ are 
\bea 
E_{\bf q,p,k}=r^2+q_0^2 +p_{1}^2\cos^2\frac{\psi_3}{2}
+2\sum_{a=1,3,4} U_a(k_a+\frac{1}{2}),
\eea
with $q_0$ and $p_1$ as eigenvalues of $i\partial_t$ and
$P_{2}$ respectively, and $k_a$ as
harmonic oscillator numbers.

The one-loop effective action is given by (16) and using 
\footnote{Here equality is exact in contrast to (17) because  
at least one of $P_i$'s is zero.}
\bea
-\ln\det^\haf(\partial_t+\sum_{i=1}^{9}P_i\gamma_i)=
-\frac{1}{4}\tr\ln(P_\lambda^2+\frac{i}{2}F_{ij}\gamma_{ij}),
\eea
one can calculate $W$ 
\bea
W\sim \int dt 
\frac{\sum_{a=1}^3\cos(2\psi_a)-4\prod_{a=1}^3\cos\psi_a }
{\cos\frac{\psi_3}{2}\prod_{a=1}^3\sin\psi_a}   \ln|r|
\eea
which is in agreement with string theory results \cite{SJ}.


{\bf 3) Two angles}

By putting $\psi_4=\psi_3=0$ in (8) the eigenvalues of $P^2_\lambda$ are
($2l=p$)
\bea\label{2.101} 
E_{\bf q,p,k}=r^2+q_0^2 +\sum_{i=1}^{l-1}(q_i^2+p_i^2)  
+2\sum_{a=3,4} U_a(k_a+\frac{1}{2}).
\eea
where $q$ and $p$'s are the eigenvalues of those $P$'s 
which do not behave as harmonic 
oscillators. After calculations similar to the above cases, 
for large separation between branes we find: 
\bea\label{2.116} 
W\sim 
\frac{(\cos\psi_1-\cos\psi_2)^2}{\sin\psi_2\sin\psi_1} 
\left(\frac{1}{r^{5-p}}\right)+\cdot\cdot\cdot ,
\eea
which is in agreement with string theory results \cite{SJ}.

The above interaction vanishes in $\psi_2=\psi_1$ cases, signalling 
enhancement of SUSY \cite{SJ,TI}.

{\bf 4) One angle}

By putting $\psi_4=\psi_3=\psi_2=0$ one finds the eigenvalues 
of $P^2_\lambda$ as 
\bea 
E_{{\bf q},{\bf p},k}=r^2+q_0^2+\sum_{i=1}^{l-1}(q_i^2+p_i^2)+
\cos^2 \frac{\psi_1}{2}q_l^2 +2 U_3(k_3+\frac{1}{2}).
\eea
For large separation between branes we find: 
\bea
W\sim \tan\frac{\psi_1}{2}\,\sin^2\frac{\psi_1}{2}\,\frac 1{r^{6-p}}+
\cdot\cdot\cdot ,
\eea
again in agreement with string theory results \cite{SJ,ArS}.
         
\noindent{\large\bf Acknowledgement}
\medskip

\noindent 
It is a pleasure to thank  M.M. Sheikh-Jabbari for useful discussions. 

\vspace{.3cm}
{\Large{\bf Appendix}}

Here we calculate the trace appeared in (\ref{2.55}).
$$ 
Tr\bigg(\frac{1}{\partial_t\gamma_9+\sum_{i=1}^{8}P_i\gamma_9
\gamma_i}\bigg)=
Tr\bigg(\frac{\gamma_9}{\partial_t+\sum_{i=1}^{8}P_i\gamma_i}\bigg).
$$
Multiplying the denominator and the nominator 
by $-\partial_t+\sum_{i=1}^{8}P_i\gamma_i$ and
using Schwinger's time representation one finds
$$
Tr\bigg(\frac{\gamma_9}{\partial_t+\sum_{i=1}^{8}P_i\gamma_i}\bigg)=
Tr\bigg(\gamma_9(-\partial_t+\sum_{i=1}^{8}P_i\gamma_i)\times
\int_0^\infty ds \exp\{ -s[-\partial_t^2+P_i^2+
\frac{i}{2}\sum_{i,j=1}^8 F_{ij}\gamma_{ij}]\}\bigg).
$$
The term proportional to $-\partial_t$ 
vanishes because it is an odd function 
with respect to the eigenvalues of $\partial_t$ 
in the trace. The next 8 terms containing $\gamma_i$'s give
zero because they are 
multiplications of odd numbers of $\gamma$ matrices
in the even number of $\gamma$ of exponential, by knowing 
$\gamma_9=\gamma_1\gamma_2\gamma_3\gamma_4\gamma_5\gamma_6\gamma_7
\gamma_8$.


\end{document}